\documentclass{article}

\usepackage{arxiv}

\usepackage[utf8]{inputenc} 
\usepackage[T1]{fontenc}    
\usepackage{hyperref}       
\usepackage{url}            
\usepackage{booktabs}       
\usepackage{amsfonts}       
\usepackage{nicefrac}       
\usepackage{microtype}      
\usepackage{lipsum}

\usepackage{amsmath}
\usepackage{amssymb}
\usepackage{amsthm}
\usepackage{epsfig}
\usepackage{graphicx}
\usepackage{color}
\usepackage[normalem]{ulem} 
\usepackage{makeidx}
\usepackage{xspace}
\usepackage{wrapfig}
\usepackage{floatrow}
\usepackage{caption,subcaption}
\usepackage{float}
\usepackage{url}
\usepackage[numbers,sort]{natbib}
\usepackage[colorinlistoftodos]{todonotes}
\usepackage{xspace}
\makeindex

\makeatletter
\newcommand{\printfnsymbol}[1]{%
  \textsuperscript{\@fnsymbol{#1}}%
}
\makeatother

\author{
  Dmitry Gordeev\thanks{All authors contributed equally.} \\
  H2O.ai\\
  \texttt{dmitry.gordeev@h2o.ai} \\
   \And
Philipp Singer\printfnsymbol{1} \\
 H2O.ai\\
  \texttt{philipp.singer@h2o.ai} \\
     \And
Marios Michailidis\printfnsymbol{1}\\
 H2O.ai\\
  \texttt{marios.michailidis@h2o.ai} \\
       \AND
Mathias Müller\\
 H2O.ai\\
  \texttt{mathias.mueller@h2o.ai} \\
      \And
SriSatish Ambati\\
 H2O.ai\\
  \texttt{sri@h2o.ai} \\
}

\date{\vspace{-5ex}}

\begin{document}

\title{Backtesting the predictability of COVID-19}

 




\newcommand{\covid}{\mbox{COVID-19}\xspace}

\maketitle
\setcounter{page}{0}

\begin{abstract}

The advent of the \covid pandemic has instigated unprecedented changes in many countries around the globe, putting a significant burden on the health sectors, affecting the macro economic conditions, and altering social interactions amongst the population through a number of mitigation measures and governmental instructions. In response, the academic community has produced multiple forecasting models, approaches and algorithms to best predict the different indicators of \covid, such as the number of confirmed infected cases, the number of deceased and economic indicators. 
Specifically at the beginning of the pandemic, researchers had little to no historical information about the pandemic at their disposal in order to inform their forecasting methods. Our work studies the predictive performance of models at various stages of the pandemic to better understand their fundamental uncertainty and the impact of data availability on such forecasts.



We use historical data of infected, deceased and recovered cases of \covid from 253 regions from the period of 22nd January 2020 until 22nd June 2020 to predict, through a rolling window backtesting framework, the cumulative number of infected cases for the next 7 and 28 days. We implement three simple models to track the root mean squared logarithmic error in this 6-month span, a baseline model that always predicts the last known value of the cumulative confirmed cases, a power growth model and an epidemiological model called SEIRD. Within our presented backtesting framework, each model is re-fitted daily and its best parameters are obtained with information known as of that point in time (in terms of historical confirmed, deceased and recovered cases) and predictions are extended for the 7-days and 28-days forecast horizons.   

We demonstrate that prediction errors are substantially higher in early stages of the pandemic, resulting from limited data. Throughout the course of the pandemic, errors regress slowly, but steadily. The more confirmed cases a country exhibits at any point in time, the lower the error in forecasting future confirmed cases. 
Our work emphasizes the significance of having a rigorous backtesting framework to accurately assess the predictive power of such models at any point in time during the outbreak which in turn can be used to assign the right level of certainty to these forecasts and facilitate better planning.



\end{abstract}

\newpage

\section{Introduction}
\label{sec:introduction}

In the event of a pandemic outbreak, stakeholders such as politicians, pharmaceutical companies or hospitals attempt to forecast the spread of the pandemic to make informed decisions about actions and policies such as lock-downs, supply chain optimization, or, in worst case, even crucial decisions about intensive care units. However, every pandemic is unique in itself and \covid reached a magnitude and severity that has not been observed over the last decades \cite{walker2020report, baker2020unprecedented, flaxman2020report}. As a result, little historical information about similar pandemics was at our disposal at the beginning of the outbreak in order to make good estimates about the future development of the disease. This information bottleneck leads to uncertainty in forecasting methods and can be crucial in the efforts to develop new medicine, vaccines, public guidelines and other important aspects to guarantee public health and safety.

\begin{figure}[t!]
\centering
\includegraphics[width=0.9\linewidth]{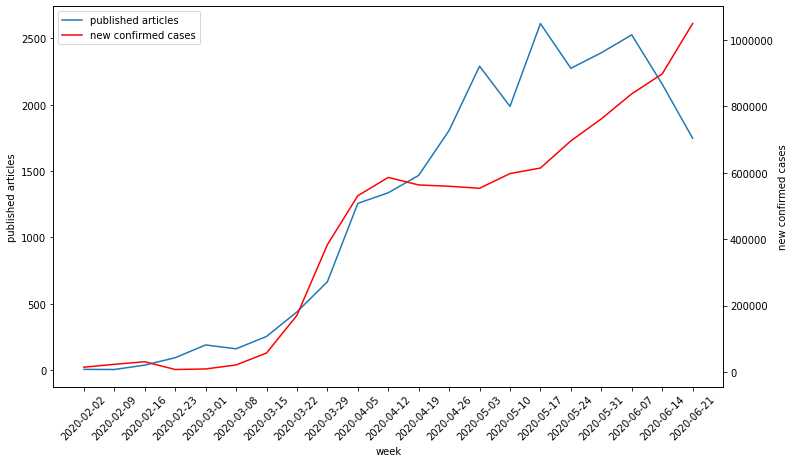}
\caption{{\bf Research publications and \covid confirmed cases.} This figure shows the number of published articles derived from the CORD-19 dataset as well as the worldwide number of new cases on a weekly aggregation level.}
\label{fig:curve}
\end{figure}

\noindent{\bf Background.}
Substantial research has been published over the course of the pandemic, as evident from the \covid Open Research Dataset (CORD-19) \cite{cord19} containing close to 30,000 \covid related research papers; the dataset has been extended to cover publications from similar corona-viruses and fostered NLP-related research on the corpus \cite{cord19kaggle, corlit}. In Figure~\ref{fig:curve}, we depict the number of research publications containing the term ``Covid`` in the title and having a publication date\footnote{All publications from WHO database do not have a date and are excluded from this visualization.} as well as the weekly number of new cases retrieved from data made available from the John's Hopkins University (see Section~\ref{sec:data}). Similar to the rapid spread of the pandemic, we observe an accelerating number of publications indicating the strong efforts of the research community to study the pandemic across disciplines.

Many different types of models have been proposed to model and forecast the number of infections within and across countries. A prominent and frequently applied type is the classical \emph{epidemiological framework} \cite{kuperman2001small} modeling susceptible, exposed, infected, and recovered agents (SEIR) that has also found its application in several \covid forecasting approaches \cite{tang2020estimation, lin2020conceptual, lopez2020modified, afonso2020epidemic, hernandez2020simple,hulshof2020not,lanl}. A second type of category represents \emph{autoregressive moving average models} that attempt to extrapolate future data by means of aggregating recent data. These types of models have had many successful implementations in time series forecasting---e.g. financial methods \cite{ariyo2014stock}---and have recently also been applied to predict \covid numbers \cite{pourghasemi2020assessment, singh2020prediction, anne2020arima}. Third, several \emph{curve fitting} and \emph{statistical} models have been proposed to be well-tailored for \covid forecasting, including power-law models \cite{ziff2020fractal}, simple linear or polynomial models \cite{yang2020early,pandey2020seir}, logistic models \cite{tatrai2020covid}, mixed-effects models \cite{ihme,ihme2}, and many others. Finally, many approaches in the realm of \emph{machine learning} have been developed \cite{covidprojections}, including e.g. Facebook's prophet algorithm \cite{ndiaye2020analysis}, gradient boosted trees \cite{suzuki2020machine}, or neural networks \cite{zhao2020well}. This list only covers a small fraction of published models, an exemplary overview of others is also given in \cite{covidhub, latif2020leveraging, currie2020simulation}.

Kaggle, a large competitive data science platform with around five million users \cite{kaggle5million}, conducted a series of five competitions \cite{kaggle1, kaggle2, kaggle3, kaggle4, kaggle5} allowing data scientists to develop and submit their \covid forecasting models to predict confirmed cases and fatalities across ${\sim}300$  regions---including mostly country-level and in certain cases province-level or state-level predictions---for no less than 30 days into the future\footnote{In each of the five competitions, at least one author of this paper finished in the top 5.}. The models were always developed on historical data and then evaluated live over a period of four succeeding weeks or more. Across all competitions, different types of models have performed well including the above-mentioned machine learning models (boosting trees, neural networks) as well as a diverse set of curve fitting, statistical, and autoregressive models. The series of competitions captures the state of development of these kinds of models during a pandemic quite well, with the models being initially quite simple and uninformed \cite{kaggle1, kaggle2}, and developing to more robust models and ensembles over time \cite{kaggle3, kaggle4, kaggle5}. While many strong solutions have been developed, it has also been shown that a lot of subjective adjustments\cite{kazanovapost, belugapost} can make a model shine or fail and that it is explicitly complex to forecast rapidly changing patterns.

\noindent{\bf Objectives.}
As summarized, a plethora of research has been conducted in order to forecast the \covid pandemic. However, given the rapidly changing environments, data irregularities as well as the inherent difficulty of predicting these numbers, this type of research has also been criticized due to the sensitivity of the topic at hand and the potential huge implications of poorly performing models \cite{jewell2020caution, failed}. Wynants et al. \cite{wynants2020prediction} conducted a review of 66 published models with focus on predicting different aspects of \covid or similar diseases including models for forecasting hospital admissions due to pneumonia, diagnostic models for detecting \covid as well as prognostic models for assessing mortality risk, length of stay in the hospital and exacerbation of the disease. Their review rated the aforementioned models as of \emph{high or unclear risk and biased due to improper testing frameworks, with non-representative selection of control patients}. \emph{They also highlighted the lack of clarity in the reporting of the findings and that these models have a high risk of overfitting}. They concluded that a reporting guideline needs to be adhered from all works predicting \covid or similar diseases to avoid unreliable predictions as the latter 
``could cause more harm than benefit in guiding clinical decisions``.

We still strongly believe in the fundamental value of these types of models, specifically for application in potential second waves or other future pandemics. In order to be able to utilize these types of models, they have to be properly evaluated and made transparent \cite{barton2020call}.

Nonetheless, most of these models have been developed during the outbreak of the pandemic, and thus, could only be evaluated on historical data up to that point. While some countries still see rising numbers in \covid infections as of this writing\footnote{Beginning of July 2020} (e.g., Brazil, India or the US), most countries are well past the peak and see rapid flattening of the curves. However a few potential instances of 2nd waves may already be happening \cite{venkatesan2020covid}. Consequently, we are now in the unique position to \emph{backtest and investigate predictive performance of \covid forecasting models across countries at different points in time.} 
This not only allows us to study the fundamental \emph{prediction difficulty of infection curves}, but also \emph{measure predictive performance at various stages of the pandemic.} 

\noindent{\bf Contributions and findings.}
To study these and similar questions, we make the following contributions: (i) We apply two simple, yet well-known and robust, \emph{short-term forecasting methods} along with a baseline to predict confirmed \covid cases. (ii) We introduce a thorough \emph{backtesting framework} that allows us to provide accurate assessments about a model's prediction performance. (iii) We utilize this framework to empirically \emph{study the general predictability of \covid across various stages of the outbreak.}

Our work highlights the importance of proper testing, tracking and quantifying of the prediction error through time as well as through different levels of accumulated infected cases. We observe that the prediction error is substantially higher in the early stages of the pandemic, when the number of confirmed cases is still low and the trends are still undeveloped. Then follows a period of approximately 15 days (past the early days of March) where the error drop significantly by about 3.5 times. From that point on it regresses steadily to lower levels as more data becomes available.      

This paper is organized as follows. Section~\ref{sec:data} describes the source of the data used as well as the methods by which the latter was transformed and processed to underpin the experiments.
A power growth model and a version of the SEIR model called SEIRD are optimized and applied via multiple moving windows in a backtesting setting across all countries to predict the confirmed infected cases of \covid and \emph{track the prediction error over time}.  Section~\ref{sec:methods} describes the methodology supporting these models in terms of their parameters, optimization routines and loss functions minimized within the context of the \emph{backtesting framework}. Section~\ref{sec:experiments} highlights the conducted experiments and core findings. Ultimately, the conclusions of the experiments are drawn in Section~\ref{sec:conclusions}.


\section{Data}
\label{sec:data}

\begin{figure}[!t]
\centering

\includegraphics[width=0.9\linewidth]{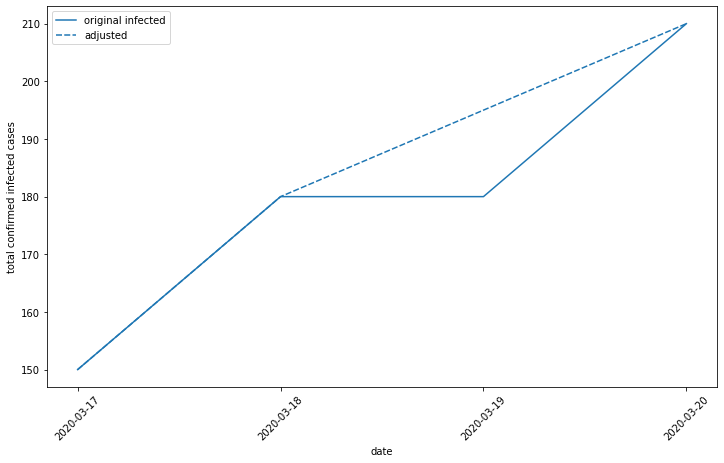}
\caption{{\bf Data preparation.} This figure depicts our data cleaning routine, where a linear interpolation is applied to replace two consecutive points of the same value.  }
\label{fig:dataprep}
\end{figure}

The primary source of data is the data repository of the Johns Hopkins University Centre for Systems Science and Engineering \cite{coviddata}. 
It contains daily updates about confirmed, deceased, and recovered cases at country level. 
Due to given irregularities in the way different countries report daily \covid statistics, we employ a basic data cleaning routine. As we are working on a cumulative level of confirmed cases, we aim at guaranteeing the monotonicity requirement. The corrective measure to ensure monotonicity is applied when the value between two consecutive dates remains the same and then increases with a high pace. In this case, the latter of the two is replaced with a linear interpolation of the neighbor values (i.e., the average between the previous value and the next value). The reasoning behind this measure is that often the cases remain the same due to irregularities and delays of the reporting system \cite{roser2020coronavirus, kucharski2020early}. An example of applying this transformation is shown in Figure~\ref{fig:dataprep}, assuming the confirmed cumulative count is originally as depicted in the solid blue line, we transform it to the dashed blue line. Even though the overall expansion of a curve may not be linear in respect to time, finding a better method to correct the same-value irregularity is an exhaustive task and out of scope for this analysis.

Overall, our dataset contains $253$ regions, with daily statistics ranging from the 22nd January 2020 to the 22nd June 2020. We observe around 9.1 million confirmed cases and $472,000$ fatalities; see also Figure~\ref{fig:curve} for a visualization of the development over time.

\section{Methodology}
\label{sec:methods}

This section details the elements utilized to implement the experiments of tracking and understanding the error in predicting the \covid confirmed infected cases over time across the globe at the country level. 
We start by elaborating our core backtesting methodology in Section~\ref{subsec:backtesting} which we utilize to study the RMSLE loss function (see Section~\ref{subsec:loss}) over time. Within the scope of backtesting, we employ three models described in Section~\ref{subsec:models}: a simple baseline model, a power growth model, and an extension of the well-known SEIR model, called SEIRD.

\subsection{Backtesting}
\label{subsec:backtesting}

Backesting---or the process of evaluating a model or an algorithm over different past periods of time---is commonly associated with trading strategies, banking, or risk prediction \cite{virdi2011review}. Backtesting in predictive modeling can be an important tool in finding the optimal parameters for the used models as well as for measuring  the volatility of predictions through time. Assessment of the forecast accuracy can be dramatically biased if done on the same data used for model fitting \cite{friedman2001elements}. A technique of setting up a single hold-out sample can serve as a way to derive more accurate forecast error estimations, however it does not provide the information about how the model accuracy improves over time, as more information becomes available. Moreover, an assessment based on a single sample is not robust given the limited size of the data available to fit the models.

In the context of the \covid pandemic, \emph{backtesting} the models aimed to predict different aspects of the disease (in confirmed, deceased, or recovered cases) can facilitate understanding of the sensitivity of these models in producing accurate and robust results given varying sizes of training history. Such an approach can enable defining the time (or the amount of training history) required to produce results of certain accuracy levels, quite often essential in order to use them efficiently in decision systems.

Using the EDI model, an exponentially decreasing intensity growth model, Moriconi \cite{moriconi2020model} used  backtesting on China’s  daily confirmed \covid cases, starting from 13th February 2020,  and observed  substantial overestimation for (approximately) the first week of predictions before the model started being significantly more accurate. 
Volatility in predictions (via varying levels of over- and underestimation) were also observed in the work of Lesage \cite {lesage2020hawkes} where the Hawkes process \cite{hawkes1971spectra} was utilized to predict via backtesting the number of confirmed cases in both France and China in different periods of February and March.

Rouabah et al. \cite{rouabah2020epidemic} used the SEIQRDP model---a variant of the SEIRD model that also incorporates quarantined (Q) individuals to be considered as active cases as well as the protected population (P) for cases that  strictly follow the standard advised protection measures---in order to forecast the elements of that model for the next six months past the last training day of 24th May 2020 for Algeria. Their work \emph{emphasizes the threat of creating unstable models due to overfitting and underfitting}, plus they point out that overfitting is a major issue in epidemic dynamical models due to the noise embedded in the data.  The SEIQRDP model’s parameters were optimized using a genetic algorithm, enhanced by published information for these parameters. To find the optimum number of iterations for the genetic algorithm to obtain the best parameters, a time-based cross validation procedure was applied in different countries so that the first $n$ days for a given country’s infected numbers are used to fit the algorithm and the last $v$ are used for validation. This process was tested on the countries of Italy, Spain, Germany and South Korea before applying it to Algeria. The ratio of $v/n$ can be adjusted based on the number of parameters that need optimization. In this case the ratio was about $1/4$. The study highlights that there is an inverse relationship between the training sample’s size and the number of iterations required in the genetic algorithm. As more data becomes available for a given country, the optimum number of iterations decreases. Therefore re-evaluating the optimization at different points in time is important for obtaining the most accurate results. 

With the application of \emph{backtesting}, we can not only derive the accuracy of predictions made in the past, but also show how the accuracy changes during the pandemic. The main idea behind backtesting is to make the predictions by the model at a fixed time point $t$ in the past and estimate the error at the time point $t+H$, where $H$ is a predefined prediction horizon. In this paper, we focus on two values, $H=7$ days for short-term predictions and $H=28$ days for a longer forecast.

Denote by $X$ a $p \times N$ matrix, where $p$ is number of regions, and $N$ is number of days with available numbers of confirmed cases. Let us denote by $X^{(t_1)(t_2)}=X_{i,j}$ where $i=1,...,p$ and $j=t_1,...,t_2$ the matrix of observations available between days $t_1$ and $t_2$ and by $X^{(t)}=X_{i,t}$ where $i=1,...,p$ the values at the day $t$. The backtesting implies fitting a model $f:\mathbb{R}^{p \times t} \rightarrow \mathbb{R}^{p \times t+H}$ for each day $t=1,...,N-H$

\begin{equation}
\hat{\theta}_t = \underset{\theta}{{\mathrm{argmin}}} \ L_{fit}(f(X^{(1)(t)};\theta)^{(1)(t)}, X^{(1)(t)})
\label{eq:btmodelfit}
\end{equation}

where $L_{fit}$ is the loss function.

We later assess the error of the forecast with horizon $H$ as

\begin{equation}
ERR_H(t)=L_{eval}(f(X^{(1)(t)};\hat{\theta}_t)^{(t+H)}, X^{(t+H)})
\label{eq:btaccuracy}
\end{equation}

where $L_{eval}$ is the evaluation metric. The experiment results in two ($H=7$ and $H=28$) time series of errors per model $f$, showing the values of forecast error, the fluctuations of the error and its dynamics during the pandemic.


\subsection{Loss function}
\label{subsec:loss}

The choice of the loss function was driven by the fact that most models predicting the development of number of confirmed infection cases assume exponential growth. In such a case, metrics like RMSE (root mean squared error) and MAE (mean absolute error), based on absolute difference in number of predicted and realized cases, tend to significantly penalize any exponential over-estimations. Therefore, root mean squared logarithmic error (RMSLE) was chosen as the evaluation metric $L_{eval}$.

\begin{equation}
RMSLE(X,Y)=\sqrt{\frac{1}{N*p} \sum_{i,j}{(ln(1+X) - ln(1+Y))^2}}
\label{eq:rmsle}
\end{equation}

where $X$ and $Y$ are matrices of the same size $N \times p$.

Also, consistent with \cite{kaggle1, kaggle2, kaggle3, kaggle4}, the loss function is applied to the cumulative number of cases.

\subsection{Models}
\label{subsec:models}

Next, we specify three models that we employ utilizing our backtesting framework to study the RMSLE error over time. (1) For reference, we utilize a simple parameter-free baseline model that predicts future confirmed cases by using the latest known data point, (2) we introduce a power-growth model employing constant growth that decays over time, and (3) we utilize a variation of the well-known epidemiological SEIR model called SEIRD.

\noindent{\bf Baseline model.}
As a reference point for the evaluation metric across the pandemic development, a simple parameter-free baseline model was applied. Denoting by $C_t$ number of cumulative confirmed cases in a region at point in time $t$, the baseline predictions are

\begin{equation}
    C_{t+i} = C_t, i=1,...,N
\label{eq:baseline_c}
\end{equation}

The baseline model is not intended to produce reasonable forecasts, but rather to indicate how difficult it is to make accurate predictions at each point in time $t$.

\noindent{\bf Power-growth model.}
This model is motivated by different types of \covid forecasting models such as statistical power law models with exponential growth \cite{mitzenmacher2004brief}, or autoregressive moving average models as described in Section~\ref{sec:introduction}. We have utilized this model successfully across all five Kaggle competitions \cite{kaggle1, kaggle2, kaggle3, kaggle4, kaggle5} and the version presented in this paper is the final adaption of it. The main idea of the model is to forecast \covid cases, by employing a constant growth rate that is derived from previous observations. This growth rate is decaying over time and the decay can accelerate. In detail, we can define the power-growth model as follows:

\begin{equation}
    C_{t+i} = C_t + gr \cdot max(0, (1+gr_d \cdot (1 + gr_{d_a})^i)^{log(t+i)})
\label{eq:powergrowth_c}
\end{equation}

\noindent
In Equation~\ref{eq:powergrowth_c} we want to predict the cumulative number of cases $C$ at time $t+i$, $i=1,...,H$ using the number of cases at time $t$; $gr$ refers to the growth rate, $gr_d$ to the decay of the growth rate, and $gr_{d_a}$ to the acceleration of the decay.

The growth rate is calculated for each region separately by taking an exponential weighted average of the observed daily growth rate over a certain number of past days ($n_{days}$). If a region does not exceed a minimum number of cases ($min\_cases$), a default growth rate ($gr_{def}$) is employed. The growth rate decay $gr_d$ as well as its acceleration $gr_{d_a}$ are constant across all regions. All parameters except the growth rate are thus hyperparameters that are optimized based on a global metric across regions. The modified power-growth model fitting was performed the following way.

\begin{equation}
\hat{\theta}_t = \underset{\theta}{{\mathrm{argmin}}} \ L_{fit}(f(X^{(1)(t-21)};\theta)^{(t-20)(t)}, X^{(t-20)(t)})
\label{eq:btmodelfitpg}
\end{equation}

Meaning that the most recent 21 days of data were used to optimize the hyperparameters. Loss function is the same as the evaluation metric $L_{fit}=L_{eval}=RMSLE$.

\noindent{\bf SEIRD model.}
The SEIR model  belongs to a family of epidemiological models (see also Section~\ref{sec:introduction} that map the spread of an epidemic through the sequential interaction of 4 groups or states (represented as ordinary differential equations), the \emph{susceptible} (or number of individuals that can contract the disease), \emph{exposed}, \emph{infected} and \emph{removed}.

Our implementation uses a variation of the SEIR model called SEIRD \cite{seird1}. In this application, the \emph{removed} category is further divided into \emph{recovered} and \emph{deceased}.  
The equations that map the rate of change in respect to the main states are displayed below:

\begin{equation}
\frac{\partial S}{\partial t} = -\beta I \frac{S}{N}
\label{eq:susceptible}
\end{equation}

where $\frac{\partial S}{\partial t}$ represents the change applied to the susceptible population $S$ at time $t$, $\beta$ is the infection rate (or how many people an infected individual infects), $I$ the infected population at time $t$ and $N$ the total population. 

\begin{equation}
\frac{\partial E}{\partial t} = \beta I \frac{S}{N} - \delta E
\label{eq:exposed}
\end{equation}

where $\frac{\partial E}{\partial t}$ represents the change applied to the exposed population $E$ at time t, $\delta$ is a parameter that controls the rate by which the exposed population transitions to the infected state and it can be interpreted as $1$ divided by the incubation period (or in other words, the period that an individual is infected but asymptomatic and unable to spread the disease to others). 

\begin{equation}
\frac{\partial I}{\partial t} = \delta E - (1 - \alpha ) \gamma I - \alpha \rho  I
\label{eq:infected}
\end{equation}

Similarly, to compute the change applied to the infected group at time $t$, a parameter $\gamma$ is used to represent the recovery rate, or how quickly individuals move to the recovered state. Equivalently, $\rho$ controls how quickly individuals move to the deceased state. The parameter $\alpha$ represents the fatality rate or the proportion of the infected population that will transition to the deceased state. The $(1-\alpha)$ represents the proportion of the infected population that will transition to the recovered state.

\begin{equation}
\frac{\partial R}{\partial t} =  (1 - \alpha ) \gamma I 
\label{eq:recovered}
\end{equation}

which expresses the change applied to the recovered group $R$ at time $t$.
Finally, 

\begin{equation}
\frac{\partial D}{\partial t} =  \alpha \rho  I
\label{eq:deceased}
\end{equation}
expresses the change applied to the deceased group $D$ at time $t$.

For each country, and given a set of bounds for the model’s parameters (of $N, \beta, \delta, \gamma, \alpha, \rho$),\emph{ a stochastic, population-based optimisation algorithm with differential evolution} \cite{storn1997differential} is applied to find the optimum values for these parameters in order to minimize the RMSE \footnote{Experiments indicated better results and faster convergence with RMSE instead of RMSLE for SEIRD; we still report RMSLE from here onwards for fair comparison.} across all the infected, recovered and deceased groups up to a selected point in time. The bounds-based optimization algorithm was preferred over others (like gradient-based), because the bounds were selected based on latest known information in regards to infection rate, incubation period, and fatality rate and it provided a fairly narrow constrained environment for the algorithm to converge more quickly. 



\begin{equation}
rmse(y, \hat{y}) = \sqrt{\frac{1}{n} \sum_{i=1}^{n}(y_{i} - \hat{y}_{i})^{2}}
\label{eq:rmseformula}
\end{equation}

Where $y$ is the observed value for either one of infected, recovered or deceased and $\hat{y}$ the corresponding predicted value for these groups. Then the overall metric to optimize can be defined as:
\begin{equation}
\widehat{N},\widehat{\beta},\widehat{\delta} ,\widehat{\gamma},\widehat{\alpha},\widehat{\rho} = argmin_{(N,\beta,\delta,\gamma,\alpha,\rho)} \  M(I,\hat{I},R,\hat{R},D,\hat{D})=\frac{rmse(I,\hat{I})+rmse(R,\hat{R})+rmse(D,\hat{D})}{3} 
\label{eq:objectiveseird}
\end{equation}

Where $M$ is the objective to minimize and connotes the average of $rmse(I,\hat{I}),rmse(R,\hat{R}),rmse(D,\hat{D})$ which are the respective root mean squared errors for infected, recovered and deceased. 
Once the optimum parameters $\widehat{N},\widehat{\beta},\widehat{\delta} ,\widehat{\gamma},\widehat{\alpha},\widehat{\rho}$ have been obtained, the curves for infected, recovered and deceased are extrapolated in time to match the forecasting period. Since the predicted values are based on the fit with the known values, it is possible that the cumulative predicted numbers are lower than the last known value. In that case, the differences between the predicted points are computed and added to the last known value to form the new cumulative predictions. 

\begin{figure}[H]
\centering
\begin{subfigure}[c]{0.95\textwidth}
\vspace{0.2em}
\includegraphics[width=0.99\linewidth]{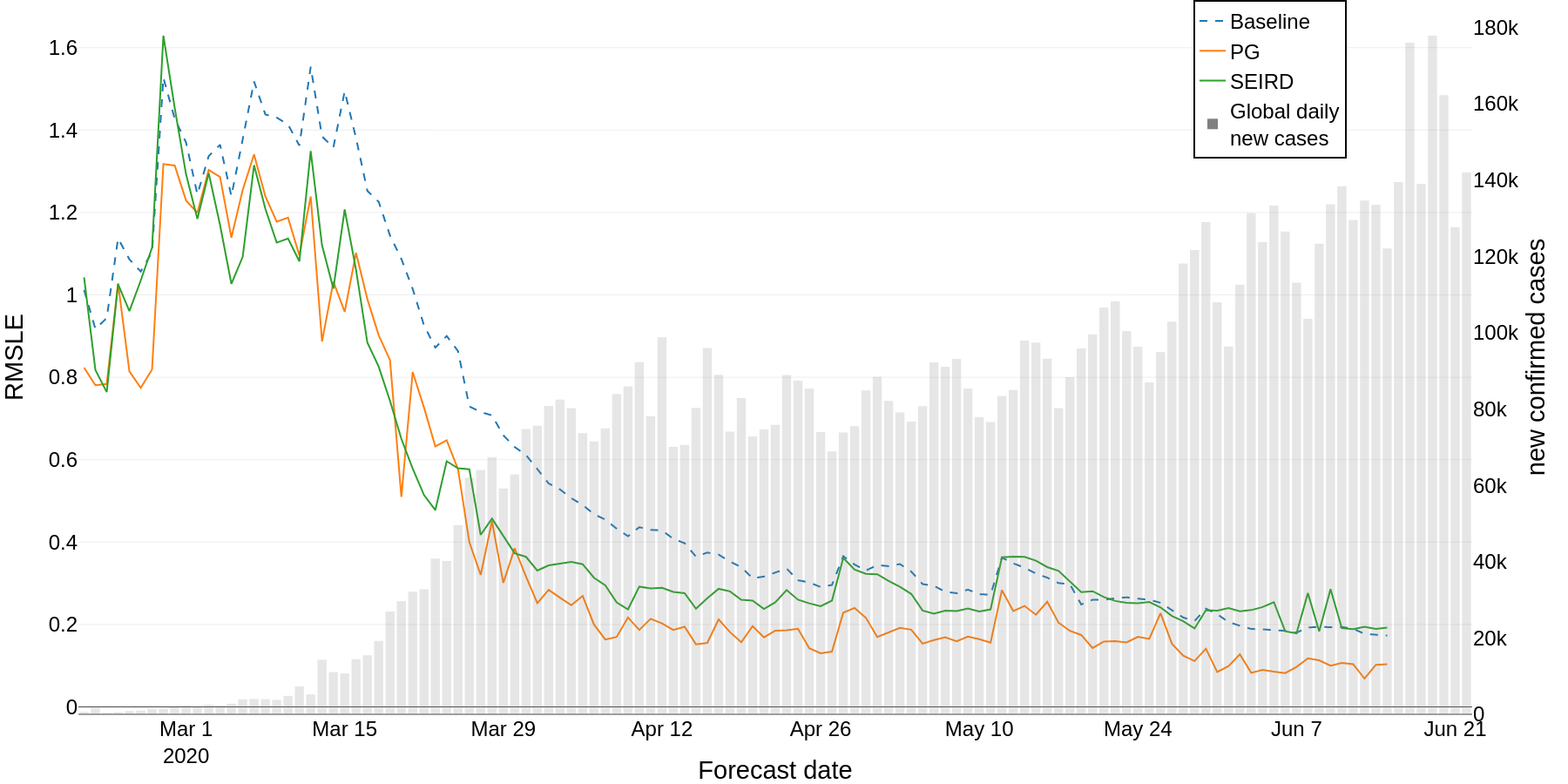}
\caption{7 days forecast horizon\vspace{2em}}
\label{subfig:backtesting7}
\end{subfigure}

\begin{subfigure}[c]{0.95\textwidth}
\includegraphics[width=0.99\linewidth]{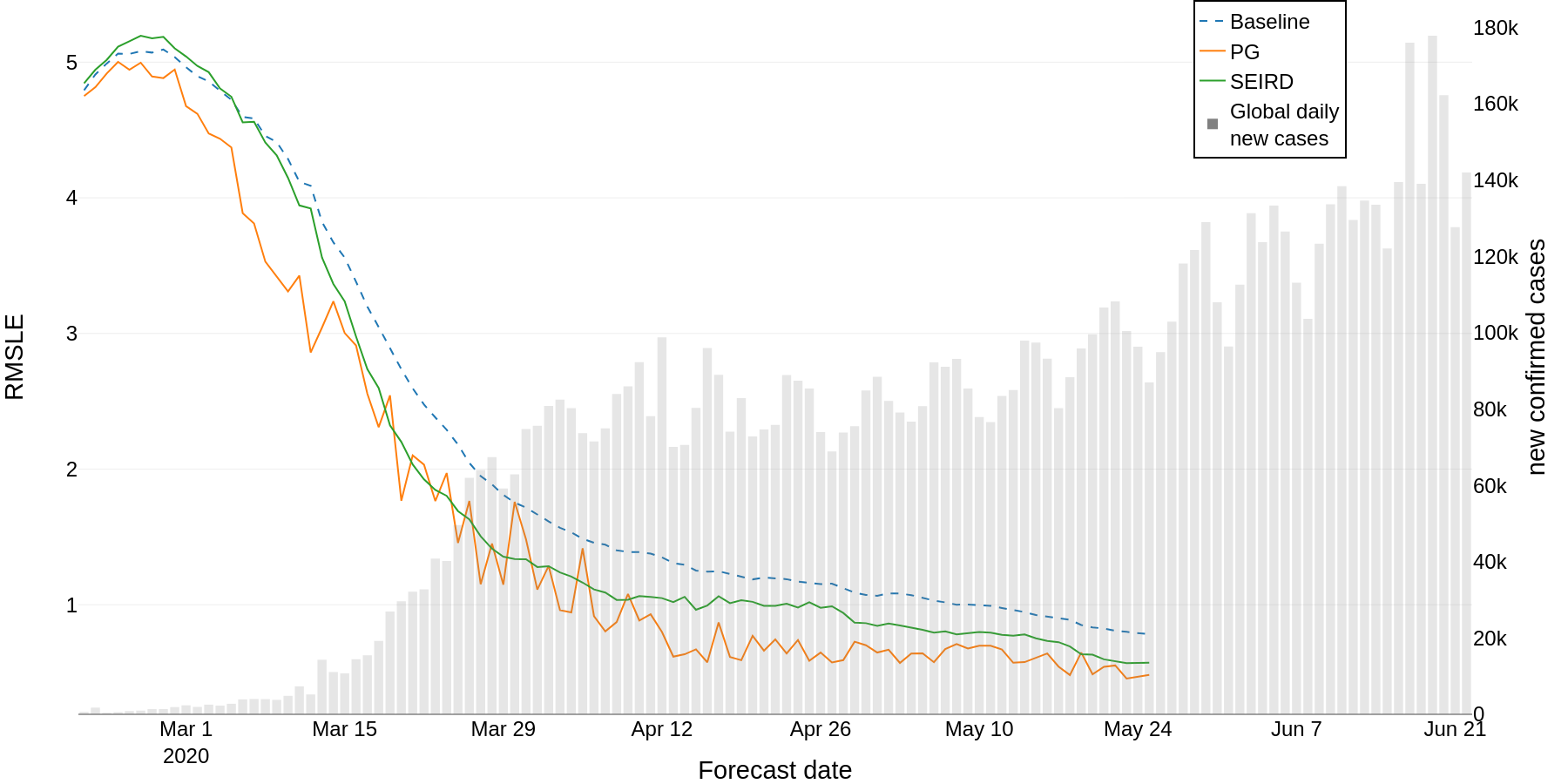}
\caption{28 days forecast horizon}
\label{subfig:backtesting28}
\end{subfigure}
\caption{{\bf Backtesting error over time.} This plot shows the errors for each model versus time (baseline, power-growth, SEIRD) for the two forecast horizons of 7 days in (a) and 28 days in (b),  limiting the analysis to regions having at least 100 confirmed cases. The x-axis depicts the forecast date---i.e., the point in time where respective model has been fitted on past data only---and the left y-axis its respective prediction error based on the given forecast horizon. For instance, in (a), the green point at May 10 refers to the RMSLE error seven days in the future for a SEIRD model fitted on all historic up to this point in time. The grey bars highlight the global number of new confirmed cases (right y-axis) on each day for visual reference. Note that the last data points plotted for (a) and (b) are 7 and 28 days prior to the last evaluation date of 22nd June 2020 as such were the periods required to calculate the evaluation error. }
\label{fig:backtesting1}

\end{figure}

\newpage
\section{Experiments}
\label{sec:experiments}

Our experiments are based on the data spanning the period from the 22nd January ($d=1$) until the 22nd June 2020 ($d=153$), counting $N=153$ days of data points from each of $p=253$ regions. In order to provide at least a month of data for training the models, backtesting results are reported from $d=31$ onwards. Two prediction horizons were chosen for the experiments: $H=7$ and $H=28$. Many regions report new cases with weekly cycle, where lower cases are reported during the weekend, therefore, horizons over full weeks are suggested to avoid instability. We make the backtesting framework as well as further code to run the experiments available online\footnote{\url{https://github.com/h2oai/covid19-backtesting-publication}}.

The first experiment aggregates the $ERR_H(t)$ by the date $t$ in order to show how forecasting error develops over the course of the pandemic. We show respective results for both forecasting horizons in Figure~\ref{fig:backtesting1}. A first observation is that it is easier to capture short-term trends, compared to long-term trends, as evident from smaller absolute prediction errors across all models for the 7-days forecasts in Figure~\ref{subfig:backtesting7} compared to the 28-days forecasts in Figure~\ref{subfig:backtesting28}. Both the power-growth and SEIRD model perform better than the simple baseline for most parts of the curve, which is why we focus on them next.

We observe a steady trend of prediction error decreasing together with error of a baseline model. We see that in the beginning of the outbreak in early March, both models depict high errors in predicting confirmed infected cases which are near the levels of 1.3 and 5 respectively for the 7-days and 28-days forecast horizons. Over time, the models’ errors move down elastically and reach 0.4 and 1.5 respectively for the two forecast horizons. In other words, the error gets reduced about three times by middle-to-end of March, which is roughly 15 extra days of observed data. From that point on the errors regress more inelastically through time and gradually reach 0.1-0.2 and 0.5 for the two forecast periods in June. 

\begin{figure}[h!t!]
\centering
\includegraphics[width=0.9\textwidth]{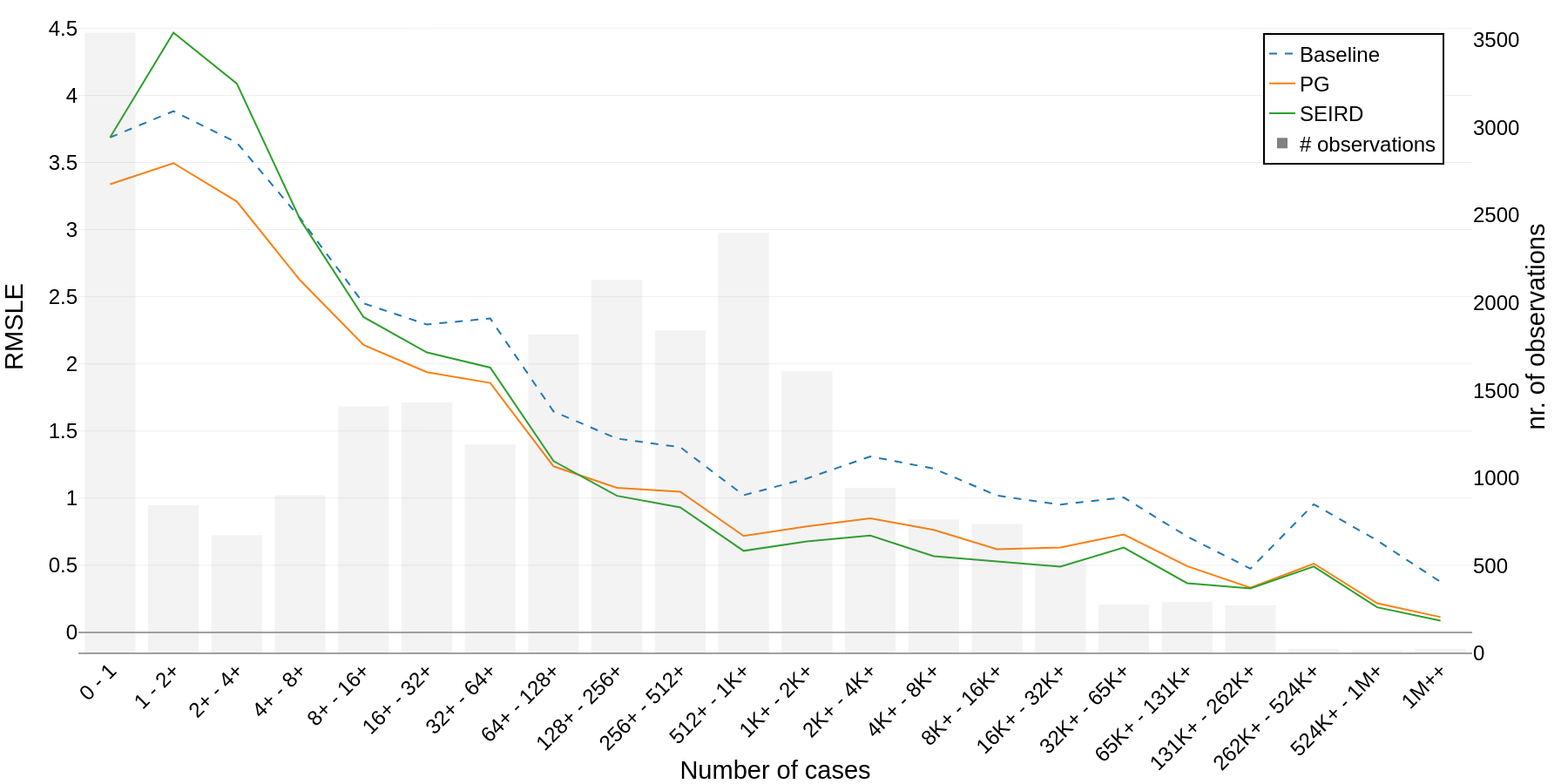}
\caption{{\bf Backtesting error by number of cases.} This plot shows how the error evolves based on the number of historical confirmed cases; it highlights the errors grouped by the number of confirmed cases in a country or region. The x-axis depicts buckets of number of cases at the point in time when the model was fitted. The left y-axis depicts the average error 28 days in the future, across all countries falling into the respective bucket of cases at any point in time. 
The grey bars (right y-axis) show how many overall observations (region-day pairs) are accumulated for each bucket. For example, the 256-512 bucket captures all observations where the historical number of confirmed cases falls into this range at any specific day of interest for fitting the models. The y-axis then depicts the average error across all region-day pairs (which sums to approximately 1,700 observations for this bucket). A single country can fall multiple times into the same bucket. 
 }
\label{fig:backtesting2}
\end{figure}

Given the decreasing error over time, we are now interested in studying the effect of historical data volume on prediction errors.  To that end, our second experiment visualized in Figure~\ref{fig:backtesting2}, contrasts how the error depends on the accumulated number of confirmed cases. The evaluation metric was aggregated by $C_t$---number of confirmed cases at the date when the forecast was made. We focus on a forecasting horizon of 28 days, but can observe similar trends for the 7-days forecast horizon. We can clearly see, that the error decreases with more training data available. SEIRD is even performing worse than the baseline with very limited number of recorded cases. As soon as a region reaches $1000$ confirmed cases, we observe that the forecast accuracy is securely below the baseline and monotonically decreasing. The power-growth model shows constant improvement of the error with increasing $C_t$.

\section{Conclusions}
\label{sec:conclusions}

In this paper, we studied the predictive performance of \covid forecasting models throughout the course of the pandemic. To that end, we examined the error (through RMSLE) for predicting \covid confirmed cases across multiple countries around the globe  \emph{through time and volume} starting from 22nd January until 22nd June 2020. The error was investigated via applying three models, a simple baseline model, a power growth model, and the well-known epidemiological SEIRD model, under a rigorous back testing framework that required refitting the models’ parameters every day on historical data up to this point and making predictions, covering the whole six-month period. We used 7-days and 28-days forecast horizons for our measurements.




Our work  \emph{highlights the importance of applying a rigorous backtesting framework} to predicting the different stages of \covid. It is clearly demonstrated (and expected) that  \emph{different time and volume can result in different error levels}. In the early days of the outbreak, when the volume of observed cases is still low, the error is larger (with higher volatility) versus later stages when the curves have more developed shapes. Accurately depicting the error level can facilitate better usage of such epidemic models when they get integrated into decision systems as it can help the decision maker decide how much confidence to place in such models at different stages throughout the epidemic. It is imperative to understand whether the error level of an epidemic model is low enough at any given point in time to provide a useful or exploitable prediction as the cost the of the models’ errors may result in more than financial losses.

{\bf Acknowledgements.} We want to thank Dr. Christof Henkel (\url{kaggle.com/christofhenkel}) for collaboration on Kaggle developing our proposed power-growth model.

\newpage
\bibliographystyle{abbrv}
{\footnotesize
\bibliography{ref}}

\end{document}